\documentclass[preprint,11pt]{article}

\usepackage{amssymb}
\usepackage{amsmath}
\usepackage{mathrsfs}
\usepackage{graphicx}
\usepackage{float}
\usepackage[version=3]{mhchem}
\usepackage[space]{grffile}
\usepackage{epstopdf}

\usepackage{authblk}

\makeatletter

\newcommand{\Rmnum}[1]{\expandafter\@slowromancap\romannumeral #1@}
\makeatother

\title{The Effect of Noise on the Propagating Speed of Pre-mixed Laminar Flame Fronts}

\author[1]{Hongliang Liu \thanks{hl854@nyu.edu}}
\author[2]{Jonathan Goodman \thanks{goodman@cims.nyu.edu}}
\affil[1]{Physics Department, New York University, NY, USA}
\affil[2]{Courant Institute of Mathematical Sciences, New York University, NY, USA}

\begin{document}

\maketitle

\begin{abstract}

We study the effect of thermal noise on the propagation speed of a planar flame.
We show that this out of equilibrium greatly amplifies the effect of thermal noise
to yield macroscopic reductions in the flame speed over what is predicted by the
noise-free model.
Computations show that noise slows the flame significantly.
The flame is modeled using Navier Stokes equations with appropriate diffusive transport
terms and chemical kinetic mechanism of hydrogen/oxygen.
Thermal noise is modeled within the continuum framework using a system of stochastic partial 
differential equations, with transport noise from fluctuating hydrodynamics and reaction noise
from a poisson model.
We use a full chemical kinetics model in order to get quantitatively meaningful results.
We compute steady and dynamic flames using an operator split finite volume scheme.
New characteristic boundary conditions avoid non-physical boundary layers at computational
boundaries.
New limiters prevent stochastic terms from introducing non-physical negative concentrations.
This represents the first computation of a model with thermal noise is a system with this
degree of physical detail.

\end{abstract}


\section{Introduction}

This paper studies the effect of thermal noise on the speed of propagating flames.
This is interesting for several reasons, even though it may seem that thermal noise would
have a small impact on something as big as a laboratory flame.
One reason is that even ``steady'' frames are far from thermodynamic equilibrium.
It is documented in other systems that out of equilibrium systems can amplify thermal
noise many orders of magnitude (e.g., \cite{Giant Fluc 0}, \cite{Giant Fluc 1}).
Perturbation analysis suggests that thermal noise of size $\varepsilon$ can change the 
speed of a propagating front by order $\varepsilon^2$.  
But detailed studies of a model system show that the effect can be much larger, on the
order of $\log^{-2}(\varepsilon)$ \cite{Noise on traveling wave} \cite{speed shifting due to a cutoff}.
It is worth noting that the physical mechanism for slowing traveling 
waves \cite{speed shifting due to a cutoff} (quenching at the leading edge) is different
from the physical mechanism of the giant fluctuations \cite{Giant Fluc 1} in phase separation
experiments (large concentration gradients).
A flame has both these ingredients, an important leading edge and large gradients.

Steady propagating flames are interesting from various scientific points of view in addition
to their obvious interest in engineering.
Several of the many physical processes involved are not quantitatively understood and are
hard to measure directly.
Examples include gradient and composition dependent transport coefficients and third body
terms in exothermal reactions.
An example of the latter is the reaction $\mbox{H} + \mbox{OH} = \mbox{H}_2\mbox{O}$ 
(it is traditional in the combustion 
literature to use $=$ instead of $\rightleftharpoons$).
The $\mbox{H}_2\mbox{O}$ will break quickly if it cannot transfer energy to a third body.
These difficult to model coefficients may be inferred from more complex physical 
experiments such as observations of
steady flame speeds (e.g., \cite{Book: Combusiton Physics}).

Using flames to calibrate the physical processes involved requires an accurate quantitative 
model of flames.
All commonly used calibrations (see, e.g., \cite{FlameSampling}) use deterministic
models of chemistry and fluid mechanics.
These models will lead to inaccurate calibrations if thermal noise has even a moderate
impact on flame speeds.
The flames that are most important in many applications are also the most sensitive 
to modeling errors.
These are ``lean'' flames, which means that the ratio of $\mbox{H}_2$ to $\mbox{O}_2$ 
is such that a large fraction of the $\mbox{O}_2$ is not consumed and the resulting flame
is not as hot (so it produces less $\mbox{NO}_x$. J. Bell, personal communication).

Although many simulations and theoretical studies have been performed exploring the
effect of thermal noise in fluids and reactions (e.g., \cite{pushedWave}, \cite{solvableModel}), none has considered as
detailed physical picture as the hydrogen flame.
As simple as it may seem, it involves a complex collection of chemical and physical processes,
not of all have agreed upon mathematical models.
The complexity seems essential, as simpler systems often have different behavior.
For example, flames behave differently if one models the chemistry as a one step reaction
of the form: $\mbox{fuel} + \mbox{O}_2 \longrightarrow \mbox{products} + \mbox{heat}$.
Many interesting phenomena do not happen without more detailed chemistry of the kind
used here (see, e.g., \cite{Book: Combusiton Physics}).
Transport processes in flames are equally subtle (e.g., \cite{Book: Giovangigli}).

Most uncertain is the modeling of thermal noise in transport processes.
A good model for simple fluids at rest is the {\em fluctuating hydrodynamics} model
of Landau and Lifschitz \cite{Book: LL Fluid}.
This model is based on detailed balance/fluctuation-dissipation arguments.
It is in some sense mesoscopic.
It uses a continuum description of the fluid and adds in thermal noise through
randomness in transport currents.
It leads to noise models that are called {\em additive noise} in the theory of stochastic
processes (noise coefficients independent of the state of the system).
With additive noise, there is no distinction between Ito and Stratonovich stochastic calculus.
But in a mathematical model of flames, noise coefficients are temperature and concentration dependent, which leads to
{\em multiplicative noise}.
The mathematical models advocated by \"{O}ttinger \cite{Book: ottinger} either are not
practical or not accepted for actual simulations of fluids with thermal noise 
(e.g., \cite{Alejandro Bell: Fluct Hydro Monte Carlo}).

In this paper, we use the generalized fluctuating hydrodynamics model advocated by Landau and Lifshitz (e.g., \cite{fluctating hydrodynamics}) to allow multiplicative noise, and use detail reactions in hydrogen flames. Except thermal noise in transport processes, we also considered noise in chemical reactions.
To integrate fluctuating governing equations, we develop a robust first-order numeric method, which is based on the operator splitting. The governing equations are split to five parts, hyperbolic flux, diffusion flux, stochastic flux, deterministic reaction, and stochastic reaction terms. We use a finite volume representation to do the spatial discretization and then use a forward Euler algorithm to advance the solution. 

At boundaries, we develop a characteristic boundary condition (extended NSCBC) similar to the one developed by \cite{Boundary Conditions 2} \cite{Boundary Conditions 1}. Initial values are specified by an approximated stable profile by a continuation method. Fluctuations in hydrogen flames can easily cause non-physical values in the numerical calculation, so limiters are introduced. We use limiters for both fluctuations in transport processes and in chemical reactions.

Applying our numerical method to $H_2$/air pre-mixed laminar flames, by the example with the equivalence ratio $\phi = 0.63$, we find that noise in the transport processes and noise in the reaction rates both slow down the flame front propagating speed. 

In following sections, in Sec. \ref{sec: dynamics}, we summarize the governing equations for multicomponent reacting flows; in Sec. \ref{sec: Numerical Methods}, we describe the numerical method used to simulate laminar flames; in Sec. \ref{sec: example}, we study the hydrogen planar flame; and in the last Section, we do a further discussion.

\section{Dynamics} \label{sec: dynamics}
We consider the hydrogen/air combustion system as a fluid mixture. Its dynamics is governed by the conservation laws of mass momentum, energy of the mixture, and the number of atoms \cite{Book: Combusiton Physics} \cite{Book: Giovangigli} \cite{Alejandro Bell: Fluct Hydro Monte Carlo}. While ignoring gravity, the governing equations can be written as,
\begin{eqnarray} 
	\frac{\partial \rho}{\partial t} + \nabla \cdot (\rho \mathbf{v}) &=& 0,  \label{eqn: conserve rho}
	\\
	\frac{\partial (\rho Y_i)}{\partial t} + \nabla \cdot (\rho Y_i \mathbf{v}) + \nabla \cdot (\rho Y_i \mathbf{V}_i) &=& \omega_i,  \quad i = 1, \cdots, N_s \label{eqn: species i}
	\\
	\frac{\partial (\rho \mathbf{v})}{\partial t} + \nabla \cdot (\rho \mathbf{v} \otimes \mathbf{v}) + \nabla \cdot \mathbf{P} &=& 0, 
	\\
	\frac{\partial (\rho E)}{\partial t} + \nabla \cdot (\rho \mathbf{v}E) + \nabla \cdot \mathbf{q} +\nabla \cdot(\mathbf{v \cdot P})  &=&   0,  \label{eqn: conserve E}
\end{eqnarray}
where $\rho$, $\mathbf{v}$, $E$, and $N_s$ denote the density, fluid velocity, specific total energy, and total number of species, respectively, of the mixture. $Y_i$ is the mass fraction of species $i$. $\mathbf{P}$, and $\mathbf{q}$ denote the pressure tensor and heat flux vector.
$\mathbf{V}_i$, and $\omega_i$ denote the diffusion velocity and mass reaction rate of species $i$, respectively.

To close the above equation system, we need to model the relationship between our chosen primitive variables and $\mathbf{V}_i$, $\omega_i$, $\mathbf{P}$,  $E$, $\mathbf{q}$ . We describe it under the following approximations: ignoring Soret, Dufour, and radiative transport processes, and  treating the fluid as a mixture of perfect gases.

The specific total energy $E$ is given by
\begin{equation}
	E = \frac{1}{2} \lvert \mathbf{v} \rvert ^2 + \sum_{i=1}^{N_s}Y_i e_i(T), \qquad \text{and} \qquad e_i(T) = e_i(T_0)+C_{v,i} (T-T_0), 
\end{equation}
where $e_i(T)$, $e_i(T_0)$ are the specific internal energy of species $i$ at temperature $T$ and at reference temperature $T_0$, , $C_{v,i}$ is heat capacity at constant volume of species $i$.

The pressure tensor is
\begin{equation}
	\mathbf{P} = p \mathbf{I} + \mathbf{\Pi} + \widetilde{\mathbf{\Pi}},
\end{equation}
where $p$ is the thermal pressure, $\mathbf{I}$ is the unit tensor, $\mathbf{\Pi}$ is the viscosity stress tensor, and $\widetilde{\mathbf{\Pi}}$ is the corresponding fluctuation term. For an ideal gas mixture, $p$ is related to the temperature $T$ by the equation of state,
\begin{equation}
	p = \rho R_u T \sum_{i=1}^{N_s}\frac{Y_i}{W_i} \; .
\end{equation}
The universal gas constant is $R_u = k_B N_A$, where $k_B$ is Boltzmann's constant and $N_A$ is Avogadro's number. The molecular weight of species $i$ is $W_i = m_i N_A$ where $m_i$ is the mass of a molecule of that species. Under the Newtonian and isotropic assumption, the viscosity stress tensor is 
\begin{equation}
	\Pi_{ij} =  - \eta[\frac{\partial v_i}{\partial x_j} + \frac{\partial v_j}{\partial x_i}] - \delta_{ij}(\zeta - \frac{2}{3}\eta)(\nabla \cdot \mathbf{v}),
\end{equation}
where $\delta_{ij}$ is the Kronecker delta, $\eta$ is the coefficient of shear viscosity, and $\zeta$ is the coefficient of bulk viscosity.

The diffusion velocity $\mathbf{V}_i$ is defined as
\begin{equation}
	\mathbf{V}_i = \mathbf{v}_i-\mathbf{v},
\end{equation}
where $\mathbf{v}_i$ is the velocity of species $i$, and $\sum \rho_i \mathbf{v}_i = \rho \mathbf{v}$.
Usually, $\mathbf{V}_i$ is written as
\begin{equation}
	\rho Y_i \mathbf{V}_i = \mathbf{F}_i + \widetilde{\mathbf{F}}_i.
\end{equation}
When the spacial gradient of pressure can be ignored and the $N_s^{th}$ species is a dominant one, we can write $\mathbf{F}_i$ as,
\begin{equation}
	\mathbf{F}_i = -\rho D_{i,N_s} \nabla Y_i = - D_i \nabla Y_i, \qquad D_i \equiv \rho D_{i,N_s}
\end{equation}
where $D_{i,N_s}$ is the diffusion coefficients of species $i$ in species $N_s$. $\widetilde{\mathbf{F}}_i$ is the corresponding fluctuation term.

The heat flux vector $\mathbf{q}$ is given by,
\begin{equation}
	\mathbf{q} = \mathbf{Q} + \mathbf{\widetilde{Q}}.
\end{equation}
Here
\begin{equation}
	\mathbf{Q} = \sum_{i=1}^{N_s}h_i \mathbf{F}_i - \lambda \nabla T,
\end{equation}  
where $\lambda$ is the thermal conductivity of the mixture, and $h_i$ is the specific enthalpy of species $i$ with $h_i = e_i + R_u T/W_i$. $\widetilde{\mathbf{Q}}$ is the corresponding fluctuation term.

The source term $\omega_i$ in Equation (\ref{eqn: species i}) is the mass rate of production of species $i$ due to chemical reactions. It also contains two parts,
\begin{equation}
	\omega_i = \Omega_i + \widetilde{\Omega}_i,
\end{equation} 
where $\Omega_i$ is the deterministic/mean rate and $\widetilde{\Omega}_i$ is its fluctuation.
We assume that the species $\{ M_1, \cdots, M_{N_s} \}$ participate in $N_r$ reactions, represented by
\begin{equation}
	\sum_{i=1}^{N_s} \nu_{i,l}' M_i 
	\;\;\stackrel{K_l}{\longrightarrow} \;\;\; \sum_{i=1}^{N_s} \nu_{i,l}'' M_i  \qquad l=1, \cdots, N_r, 
\end{equation}
where $K_l$ is the reaction rate of the $l_{th}$ reaction. The deterministic chemical reaction rates are given by
\begin{equation}
	\Omega _i = \sum\limits_{l=1}^{N_r} \nu_{i,l} W_i K_l \prod_{j=1}^{N_s}(\frac{\rho Y_j}{W_j})^{\nu_{j,l}'},
\end{equation}
where $ \nu_{i,l} = \nu_{i,l}'' - \nu_{i,l}' $ is the stoichiometric coefficient associated with species $i$ in the $l_{th}$ reaction.

We assume the fluctuations are white noise. Advocated by Landau and Lifshitz \cite{Book: LL Fluid}, we derive the magnitudes of fluctuations corresponding to transport processes by the fluctuation-dissipation theorem (see \ref{appendix 2}).

The stochastic viscous flux tensor is a Gaussian random field with covariance given by
\begin{equation}
	\langle \widetilde{\Pi}_{ij}, \widetilde{\Pi}_{kl} \rangle = [2k_B T \eta (\delta_{ik} \delta_{jl} + \delta_{il} \delta_{jk} ) + 2k_B T(\zeta - 2/3 \eta)\delta_{ij} \delta_{kl}] \delta{(t-t')} \delta{( \mathbf{r}-\mathbf{r'})}.
\end{equation}
When we choose $\rho, \mathbf{v}, E$ and $Y_i$, with ${i=1, \cdots, N_s-1}$ as independent variables, the stochastic diffusion flux of species $i$ is given by
\begin{equation}
	\widetilde{F}_{i,l} = \sqrt{ \frac{2D_i}{(\frac{N_A}{W_i} \frac{1}{Y_i} + \frac{N_A}{W_{N_s}} \frac{1}{Y_{N_s}})}} W^{(Y_i,l)},
\end{equation}
where $l$ can be 1,2,3, which stands for the $x$, $y$, or $z$ direction, $i = 1,2,\cdots, N_s-1$, and $W^{(Y_i, l)}$ is a white noise random Gaussian vector with uncorrelated components.
The stochastic heat flux is
\begin{equation}
	\widetilde{\mathbf{Q}} = \sqrt{2k_B \lambda T^2} W^{(T)} + \sum\limits_{i=1}^{N_s-1}(h_i - h_{N_s})\widetilde{\mathbf{F}}_i,
\end{equation}
where $W^{(T)}$ is a white noise random Gaussian vector with uncorrelated components.

We treat chemical reactions as independent Poisson processes. For one reaction, in a given time $\Delta t$, suppose the average of the number of happened reactions is $a \Delta t$. Then the variance of this number is $a \Delta t$. When  $a \Delta t$ is big, we can approximate a Poisson distribution as a Gaussian distribution. Following this argument, we get the expression for $\widetilde{\Omega}_i$.
\begin{equation}
	\widetilde{\Omega}_i = \frac{1}{\sqrt{N_A}}\sum\limits_{l=1}^{N_r} \nu_{i,l} W_i \sqrt{K_l \prod_{j=1}^{N_s}(\frac{\rho Y_j}{W_j})^{\nu_{j,l}'} } W_l^{(R)},
\end{equation}
where $W_l^{(R)}$ is a white noise random Gaussian vector with uncorrelated components. 

In this paper, we set the bulk viscosity $\zeta = 0$. The shear viscosity $\eta$, the thermal conductivity $\lambda$, and diffusion coefficients $D_i$ are all treated as constants. For mass conservation, we require $\sum_{i=1}^{N_s}Y_k = 1$, $\sum_{i=1}^{N_s} \Omega_i = 0$, $\sum_{i=1}^{N_s} \mathbf{F}_i = 0$, $\sum_{i=1}^{N_s} \widetilde{\Omega}_i = 0$ and $\sum_{i=1}^{N_s} \widetilde{\mathbf{F}}_i = 0$. In our research, we consider a combustion in the air ( $29\% O_2$ + $ 71\% N_2$). We put $N_2$ to be the $N_s^{th}$ species. We choose $\rho, \mathbf{v}, E$ and $Y_i$, with ${i=1, \cdots, N_s-1}$ as independent variables. The requirements for mass conservation are automatically satisfied when we set $Y_{N_s} = 1 - \sum\limits_{i=1}^{N_s-1} Y_i$. We point out here that $\widetilde{\mathbf{\Pi}}$,  $\widetilde{\mathbf{F}}_i$, and $\sqrt{2k_B \lambda T^2} W^{(T)}$ are uncorrelated and we assume that they are also uncorrelated with $\widetilde{\Omega}_i$.

\section{Numerical Methods} \label{sec: Numerical Methods}
To numerically integrate (\ref{eqn: conserve rho}) to (\ref{eqn: conserve E}), we write the governing equations in the following form,
\begin{equation}
	\frac{\partial \mathbf{U}}{\partial t} = -\nabla \cdot \mathbf{F} +\mathbf{S}, \label{eqn: governing}
\end{equation}
where $\mathbf{U} = [\rho, \rho Y_i, \rho \mathbf{v}, \rho E]$ is the set of conservative variables, $\mathbf{F}$ stands for all kinds of flux terms, and $ \mathbf{S} $ stands for source terms, i.e. reaction terms. $\mathbf{F} = \mathbf{F}_H + \mathbf{F}_D + \mathbf{F}_S$, where $ \mathbf{F}_H$, $ \mathbf{F}_D$, and $ \mathbf{F}_S$ are the hyperbolic, diffusive, and stochastic flux terms. $ \mathbf{S} = \mathbf{R}_D + \mathbf{R}_S $, where $\mathbf{R}_D$ and  $ \mathbf{R}_S$ are deterministic and stochastic parts of chemical reaction terms. 

The fluxes are given by
\begin{equation}
	\mathbf{F}_H = 	
		\left[  \begin{array}{c}
				\rho \mathbf{v} \\
				\rho \mathbf{v} Y_i \\
				\rho \mathbf{v} \mathbf{v}^T + p \mathbf{I} \\
				\rho \mathbf{v} (E+p)
				\end{array} \right]; \qquad
	\mathbf{F}_D = 
		\left[  \begin{array}{c}
				0 \\
				\mathbf{F}_i \\
				\mathbf{\Pi} \\
				\mathbf{Q} + \mathbf{\Pi \cdot \mathbf{v}}
				\end{array} \right];	\qquad
	\mathbf{F}_S = 
		\left[  \begin{array}{c}
				0 \\
				\widetilde{\mathbf{F}}_i \\
				\widetilde{\mathbf{\Pi}} \\
				\widetilde{\mathbf{Q}}+\widetilde{\mathbf{\Pi}} \cdot \mathbf{v} \\
				\end{array}	\right].
\end{equation}
Reaction terms are given by
\begin{equation}			
	\mathbf{R}_D = 
		\left[  \begin{array}{c}
				0 \\
				\Omega_i \\
				0 \\
				0
				\end{array} \right]; \qquad
	\mathbf{R}_S = 
		\left[  \begin{array}{c}
				0 \\
				\widetilde{\Omega}_i \\
				0 \\
				0
				\end{array} \right].
\end{equation}
For $N_s$ species, $i=1, \cdots, N_s-1$, and there are $N_s+d+1$ governing equations in $d$ dimensions.

In our research, we only deal with one dimensional system, so we write down the numeric methods in 1d formula. But it can be generated to 2d and 3d. In the following,  $\mathbf{U}(t, x)$ stands for the exact or analytical solution of Equation (\ref{eqn: governing}), and $\bar{\mathbf{U}}(t, x)$ stands for the approximated solution in the numerical process.

We use first order forward Euler to integrate these governing equations based on a finite volume representation \cite{Leveque Book}, which is written as
\begin{equation}
	\mathbf{U}_i^{n+1} = \mathbf{U}_i^n - \frac{\Delta t}{\Delta x}[\mathbf{F}_{i+\frac{1}{2}}^n - \mathbf{F}_{i-\frac{1}{2}}^n] + \Delta t \mathbf{S}_i^n,
\end{equation}
where the numerical flux $\mathbf{F}_{i-\frac{1}{2}}^n$ and $\mathbf{F}_{i+\frac{1}{2}}^n$ are fluxes at the $i_{th}$ cell's left and right faces, and $\mathbf{S}_i^n$ are the source terms the $i_{th}$ cell. 
Due to different properties of the flux and source terms, we do operator splitting,
\[
		\frac{\partial \mathbf{U}}{\partial t} = A_1 + A_2 + A_3 + A_4 + A_5,
\]
where $A_1 = -\nabla \cdot \mathbf{F}_H$, $A_2 = -\nabla \cdot \mathbf{F}_D$, $A_3 = -\nabla \cdot \mathbf{F}_S$,  $A_4 =  \mathbf{R}_D$, and $A_5 = \mathbf{R}_S$.

\subsection{Numerical Representation of the Flux and Source Terms}

The hyperbolic part can be solved by HLLC solver \cite{Toro Book} \cite{HLLC}. There are many algorithms to compute the wave speeds $S_L$ and $S_R$, such as the Roe average eigenvalues for the left and right non–linear waves \cite{Roe Speed} and pressure-based wave speed estimates \cite{HLLC}. In this paper, we use the speed estimates proposed by Einfeldt \cite{HLLE Speed}.

The diffusion flux $\mathbf{F}_D$ can be calculated by the second order centered difference 
\begin{equation}
	\nabla_{i+1/2} u= \frac{u_{i+1}-u_{i}}{\Delta x}.
\end{equation}

The stochastic flux $\mathbf{F}_S$ consists of $\sigma(\mathbf{U})W$ with $W$ standing for white noise.  The white noise cannot be evaluated as a point value in either space or time, but it is not a problem for a finite volume and a given time step. We can evaluate its spatio-temporal average value for the space domain $[x_{j}, x_{j+1}]$ and the time period $[t_n, t_{n+1}]$,
\begin{equation}
	W_{j+1/2}^n = \frac{1}{\Delta x \Delta t} \int_{n\Delta t}^{(n+1)\Delta t}\int_{j \Delta x}^{(j+1)\Delta x} W(x,t)dx dt 	
\end{equation}
which is a normal random variable with zero mean and variance $\frac{1}{\Delta t \Delta x}$, independent between different cells and time steps. The $\sigma(\mathbf{U})W$ term at the cell boundary, can be evaluated as
\begin{equation}
	(\sigma(\mathbf{U}) W)_{j+1/2}^n = \sigma (\mathbf{U}_{j+1/2}^n) \frac{1}{\sqrt{\Delta t \Delta x}}Z_{j+1/2}^n,
\end{equation}
where $\mathbf{U}_{j+1/2}^n$ can be interpolated by $\mathbf{U}_{j}^n$ and $\mathbf{U}_{j+1}^n$, and $Z_{j+1/2}^n$ are normal random variables with zero mean and variance one, independent between different cells and time steps.

The source terms $\mathbf{R}_D$ and $\mathbf{R}_S$ are relatively easy to evaluate.
The deterministic reaction terms are given by
\begin{equation}
	\mathbf{R}_{D,j}^n = \mathbf{R}_D( \mathbf{U}_{j}^n ).
\end{equation}
As the stochastic flux terms, the stochastic reaction terms consist of $\sigma(\mathbf{U})W$ with $W$ standing for white noise.  The white noise can be discretized using a spatio-temporal average,
\begin{equation}
	W_{j}^n = \frac{1}{\Delta x \Delta t} \int_{n\Delta t}^{(n+1)\Delta t}\int_{(j-1/2) \Delta x}^{(j+1/2)\Delta x} W(x,t)dx dt 	
\end{equation}
which is also a normal random variable with zero mean and variance $\frac{1}{\Delta t \Delta x}$, independent between different cells and time steps. So the stochastic reaction terms are formed from
\begin{equation}
	(\sigma(\mathbf{U}) W)_{j}^n = \sigma (\mathbf{U}_{j}^n) \frac{1}{\sqrt{\Delta t \Delta x}}Z_{j}^n,
\end{equation}
where $Z_{j}^n$ are normal random variables with zero mean and variance one, independent between different cells and time steps.

\subsection{Boundary Conditions}
At boundaries, if the hyperbolic terms dominate the governing equations, the relations based on characteristic lines  can be used to specify boundary conditions. We develop the extended Navier-Stokes characteristic boundary conditions which are similar to the one in \cite{Boundary Conditions 2}. As indicated by the name, it is an extension of Navier-Stokes characteristic boundary conditions \cite{Boundary Conditions 1} for multicomponent reactive flows.

As we know, the governing equations can be written in an equivalent form with primitive variables,
\begin{equation}
	\frac{\partial \tilde{\mathbf{U}}}{\partial t} + A \frac{\partial \tilde{\mathbf{U}}}{\partial x} + \mathbf{C} = 0,
\end{equation}
where primitive variables are chosen as $\tilde{\mathbf{U}} = (\rho, T, v, Y_i)$ and the $\mathbf{C}$ takes into account fluctuation terms and all others which do not involve any first derivative of the primitive variables $\tilde{\mathbf{U}}$ along the $x$ direction, i.e., the viscous, diffusive, and reactive parts. Furthermore, we can write them as following, in which characteristic waves are easily identified, 
\begin{eqnarray}
	\frac{\partial \rho}{\partial t} + \frac{\rho \bar{C_p}}{c^2}(\mathscr{L}_1 + \mathscr{L}_{N_s+2}) + \mathscr{L}_{N_s+1} +C_1 &=& 0 , \label{eqn: rho} \\
	\frac{\partial T}{\partial t} + \mathscr{L}_1 + \mathscr{L}_{N_s+2} + \sum_{i=2}^{N_s}\mathscr{L}_i +C_2 &=& 0 ,\\	
	\frac{\partial v}{\partial t} + (\mathscr{L}_{N_s+2}-\mathscr{L}_1) \frac{\bar{C_p}}{c} +C_3 &=& 0 ,\\	
	\frac{\partial Y_1}{\partial t} + \frac{W_1 W_{N_s}}{W_1 - W_{N_s}}\frac{\mathscr{L}_2}{T \overline W} + \frac{W_1 W_{N_s}}{W_1 -W_{N_s}} \frac{\mathscr{L}_{N_s+1}}{\rho \overline W} +C_4 &=&0 ,\\	
	\frac{\partial Y_j}{\partial t} + \frac{W_j W_{N_s}}{W_j-W_{N_s}}\frac{\mathscr{L}_{j+1}}{T \overline W} + C_{j+3} &=& 0, \label{eqn: Y_j}
\end{eqnarray}
and in Equation (\ref{eqn: Y_j}) $ j = 2, \cdots, N_s-1$. (We follow procedures in \cite{Boundary Conditions 2}. The result is different from \cite{Boundary Conditions 2}, because we choose $\tilde{\mathbf{U}} = (\rho, T, v, Y_i)$ with $i = 1, 2, \cdots, N_s-1$ as independent variables, and suppose there are no $y$ and $z$ components.)
In this set of equations, the $\mathscr{L}_i$ terms are wave amplitude variations and can be written as,
\begin{equation}
	\begin{aligned} 
		\mathscr{L}_1 = 
			 &(v-c) [ (\frac{1}{2} \frac{\gamma -1}{\gamma}\frac{T}{\rho})\frac{\partial \rho}{\partial x} + (\frac{1}{2} \frac{\gamma -1}{\gamma})\frac{\partial T}{\partial x} - \frac{1}{2} \frac{c}{\bar {C_p}} \frac{\partial v}{\partial x} \\
			 & + \sum_{i=1}^{N_s-1}(\frac{1}{2} \frac{\gamma-1}{\gamma} \overline{W}T(\frac{1}{W_i}-\frac{1}{W_{N_s}}) ) \frac{\partial Y_i}{\partial x}], \label{eqn: 1st characteristic}
	\end{aligned}
\end{equation}
\begin{equation}	
	\begin{aligned}
		\mathscr{L}_2 = 
			& v [ (\frac{1-\gamma}{\gamma})\frac{T}{\rho}\frac{\partial\rho}{\partial x} + \frac{1}{\gamma} \frac{\partial T}{\partial x} - \overline{W}T(\frac{1}{W_1}-\frac{1}{W_{N_s}})\frac{\partial Y_1}{\partial x}\\
			& + \sum_{i=1}^{N_s-1} \frac{1}{\gamma} \overline{W}T (\frac{1}{W_i} - \frac{1}{W_{Ns}}) \frac{\partial Y_i}{\partial x}],
	\end{aligned}
\end{equation}
\begin{equation}
	\mathscr{L}_{j+1} =  v[-\overline{W}T(\frac{1}{W_j}-\frac{1}{W_{N_s}}) \frac{\partial Y_j}{\partial x}], \label{eqn: j characteristic}
\end{equation}
\begin{equation}
	\begin{aligned}
		\mathscr{L}_{N_s+1} = 
			 & v[(\frac{\gamma-1}{\gamma})\frac{\partial \rho}{\partial x} -(\frac{\rho}{\gamma T})\frac{\partial T}{\partial x}
			 -\sum_{i=1}^{N_s-1} \frac{1}{\gamma} \overline{W}\rho (\frac{1}{W_i} - \frac{1}{W_{Ns}}) \frac{\partial Y_i}{\partial x}],
	\end{aligned}
\end{equation}
\begin{equation}
	\begin{aligned}
		\mathscr{L}_{N_s+2} = 
			& (v+c)[ (\frac{1}{2} \frac{\gamma -1}{\gamma}\frac{T}{\rho})\frac{\partial \rho}{\partial x} + (\frac{1}{2} \frac{\gamma -1}{\gamma})\frac{\partial T}{\partial x} + \frac{1}{2} \frac{c}{\bar {C_p}} \frac{\partial v}{\partial x} \\
			& + \sum_{i=1}^{N_s-1}(\frac{1}{2} \frac{\gamma-1}{\gamma} \overline{W}T(\frac{1}{W_i}-\frac{1}{W_{N_s}}) ) \frac{\partial Y_i}{\partial x}], \label{eqn: last characteristic}
	\end{aligned}	
\end{equation}
and in Equation (\ref{eqn: j characteristic}) $ j = 2, \cdots, N_s-1$. The wave speeds associated to different $\mathscr{L}_i$'s are respectively $v-c$ for $\mathscr{L}_1$  and $v$ for all $\mathscr{L}_j$'s with $j= 2$ to $(N_s+1)$, and $v+c$ for $\mathscr{L}_{N_s+2}$.

Assume the waves at the boundaries in the full governing reacting flow equations have the same amplitude as in the case of a local one-dimension inviscid (LODI) non-reacting flow. At the boundaries of the computational domain, the governing equations Equation (\ref{eqn: rho}) to Equation (\ref{eqn: Y_j}) are written under the assumption of negligible $C_i$ terms to obtain the LODI relations. These relations provide ``compatibility" conditions between the values of the $\mathscr{L}_i$ and the conditions used at the boundary. In terms of non-conservative variables, the LODI equations are,
\begin{eqnarray}
	\frac{\partial \rho}{\partial t} + \frac{\rho \bar{C_p}}{c^2}(\mathscr{L}_1 + \mathscr{L}_{N_s+2}) + \mathscr{L}_{N_s+1}  &=& 0 , \label{eqn: LOD1 rho} \\
	\frac{\partial T}{\partial t} + \mathscr{L}_1 + \mathscr{L}_{N_s+2} + \sum_{i=2}^{N_s}\mathscr{L}_i  &=& 0 ,\\	
	\frac{\partial v}{\partial t} + (\mathscr{L}_{N_s+2}-\mathscr{L}_1) \frac{\bar{C_p}}{c}  &=& 0 ,\\	
	\frac{\partial Y_1}{\partial t} + \frac{W_1 W_{N_s}}{W_1 - W_{N_s}}\frac{\mathscr{L}_2}{T \overline W} + \frac{W_1 W_{N_s}}{W_1 -W_{N_s}} \frac{\mathscr{L}_{N_s+1}}{\rho \overline W}  &=&0 ,\\	
	\frac{\partial Y_j}{\partial t} + \frac{W_j W_{N_s}}{W_j-W_{N_s}}\frac{\mathscr{L}_{j+1}}{T \overline W} &=& 0. \label{eqn: LOD1 Y_j}
\end{eqnarray}
LODI conditions can be written in the form of all variables of interest, such as pressure, total enthalpy, and so on. For example,
\begin{equation}
	\frac{\partial p}{\partial t} + \rho \bar{C_p}(\mathscr{L}_1 + \mathscr{L}_{N_s+2}) = 0. \label{eqn: p characteristic}
\end{equation}
LODI relations associated to gradients are straightforward from the definition of $\mathscr{L}_i$, for example,
\begin{equation}
	\frac{\partial T}{\partial x} = \frac{\mathscr{L}_1}{v-c} + \frac{\mathscr{L}_{N_s+2}}{v+c} + \frac{1}{v}\sum_{i=1}^{N_s-1}\mathscr{L}_{i+1}.
\end{equation}
So both Dirichlet and Neumann boundary conditions have corresponding LODI condition relating wave amplitude variations.

For premixed hydrogen/air laminar flames, it's usually associated with the subsonic inflow and outflow.  At the subsonic inflow boundary, only one characteristic wave is outgoing. So there are $N_s+1$ degrees of freedom, we can choose to specify $T, v, Y_i$, and calculate $\rho$. While at the subsonic outflow boundary, there are $N_s+1$ waves outgoing. So there is only one degree of freedom. We can choose to specify the pressure. If pressure is set to be a constant, according to (\ref{eqn: p characteristic}), $\mathscr L_1 = -\mathscr{L}_{N_s+2}$. $\mathscr{L}_j$ with $j = 2, \cdots, N_s+2$ can be decided according to equations (\ref{eqn: 1st characteristic}) to (\ref{eqn: last characteristic}). Following that, we can specify values for all primitive variables according to equations (\ref{eqn: rho}) to (\ref{eqn: Y_j}), or in the approximated case, according to LODI equations (\ref{eqn: LOD1 rho}) to (\ref{eqn: LOD1 Y_j}).

\subsection{Limiters}
When we consider all species in the whole combustion process, there are some species whose densities are close to zero at some cells. In this case, the Gaussian noise terms will cause negative mass fractions, which are non-physical. We use limiters for stochastic fluxes and for stochastic reaction terms.

For stochastic fluxes, assuming the original numerical method (an attempt step) is given by
\begin{equation}
	(\hat{\rho}_i)_j^{n+1} = (\rho_i)_j^n - \frac{\Delta t}{\Delta x} ( (F_{S,\rho_i})_{j+1/2}^n - (F_{S,\rho_i})_{j-1/2}^n),
\end{equation}
where $\rho_i = \rho Y_i$ is the density of species $i$, and $(F_{S,\rho_i})$ is the stochastic flux for $\rho_i$. If $(\hat{\rho}_i)_j^{n+1} \ge 0$, we set $(\rho_i)_j^{n+1} = (\hat{\rho}_i)_j^{n+1}$; otherwise, limiters are used to recalculate its value, the new numerical method is given by
\begin{equation}
	(\rho_i)_j^{n+1} = (\rho_i)_j^n - \frac{\Delta t}{\Delta x} ( a_{i,j+1/2}^n (F_{S,\rho_i})_{j+1/2}^n - a_{i,j-1/2}^n (F_{S,\rho_i})_{j-1/2}^n ),
\end{equation}
We use an iterative method to find these limiters. There are three possible cases that will give a negative $\rho_i$. We can set limiters respectively.
\begin{enumerate}
	\item $(F_{S,\rho_i})_{j-1/2}^n<0$ and $(F_{S,\rho_i})_{j+1/2}^n>0$, we choose limiters $ a_{i,j-1/2}^n = a_{i,j+1/2}^n = \frac{\Delta x}{\Delta t } \frac{(\rho_i)_j^n }{(F_{S,\rho_i})_{j+1/2}^n- (F_{S,\rho_i})_{j-1/2}^n}$;
	\item $(F_{S,\rho_i})_{j-1/2}^n \ge 0$ and $(F_{S,\rho_i})_{j+1/2}^n>0$, we choose limiters $ a_{i,j-1/2}^n = 1 $ and  $a_{i,j+1/2}^n = \frac{\Delta x}{\Delta t } \frac{(\rho_i)_j^n}{(F_{S,\rho_i})_{j+1/2}^n}$;
	\item $(F_{S,\rho_i})_{j-1/2}^n<0$ and $(F_{S,\rho_i})_{j+1/2}^n \le 0$, we choose limiters $ a_{i,j-1/2}^n = -  \frac{\Delta x}{\Delta t } \frac{(\rho_i)_j^n}{(F_{S,\rho_i})_{j-1/2}^n} $ and $a_{i,j+1/2}^n =1$.
\end{enumerate}
This process is repeated until there are no negative mass fractions for current time step.

For stochastic reaction terms, a similar process is applied. Assuming the original numerical method (an attempt step) is given by
\begin{equation} \label{eqn: react noise algo before limiter}
	(\hat{\rho}_i)_j^{n+1} = (\rho_i)_j^n - \Delta t (\widetilde{\Omega}_i)_j^{n}.
\end{equation}
If $(\hat{\rho}_i)_j^{n+1} \ge 0, \forall i$, we set $(\rho_i)_j^{n+1} = (\hat{\rho}_i)_j^{n+1}$; otherwise, limiters are used to set the values of $(\rho_k)_j^{n+1}$ ($k=1,\cdots, N_s-1$), the new numerical method is given by
\begin{equation}
	(\rho_k)_j^{n+1} = (\rho_k)_j^n - a_{j}^n \Delta t (\widetilde{\Omega}_k)_j^{n},
\end{equation}
where $a_{j}^n = \min_{i: (\hat{\rho}_i)_j^{n+1}<0} \{\frac{(\rho_i)_j^n }{ \Delta t (\widetilde{\Omega}_i)_j^{n}} \}$. Note that $a_{j}^n$ will be applied to species $k$, $k=1,\cdots, N_s-1$, because $(\widetilde{\Omega}_k)_j^{n}$ are related to $(\widetilde{\Omega}_i)_j^{n}$. This process is repeated until there are no negative mass fractions  for current time step.

\section{Numerical Simulation of Hydrogen/Air Flame } \label{sec: example}
We apply the above numerical method to simulate a premixed laminar hydrogen flame. We consider the burning of hydrogen in air with the equivalence ratio $ \phi <1$, which is a lean flame. We ignore other components in air except oxygen and nitrogen. In a lean flame, the temperature is not too high, so we ignore the reaction related to nitrogen. Overall we take nine species into account, $N_s = 9$, eight reactive ones and nitrogen $N_2$. When $ \phi <1$,  $N_2$ is the abundant species. So the diffusion coefficients of species $i$ is mainly governed by its interaction with nitrogen. In this paper, we have simulated $\phi = 0.63$.

\subsection{Chosen of Initial Conditions and the Inflow Speed}
As mentioned in the above section, in order to use the LODI approximation at boundaries, the viscous, diffusive, reaction and fluctuation terms should be able to be ignored. When the inflow speed is close to the flame front propagating speed (for fluctuation situations, the mean flame front propagating speed), one can choose a domain, which is much bigger than the flame thickness, and set an initial condition which is close to the profile of a well developed flame with the combustion happening in the middle. In this case, the LODI approximation at boundaries will be a good approximation for a long enough period. In order to be able to do it, one needs to estimate the flame speed and its spatial profile.

According to experimental results, the speeds of premixed laminar hydrogen flames are usually in the magnitude of meters per second, and the velocity field is generally in the same magnitude of flame speed, i.e. the Mach number $M = \frac{v}{c}$ is small. We turn off fluctuation terms and use the low Mach number approximation. By the continuation method \cite{Continuation method}, we can find the steady state solution for the low Mach number deterministic model quickly, and use it as the initial conditions and roughly estimate flame speed.

Our numerical experiment parameters are chosen as following,  the length of spatial domain is $0.1$ cm, $\Delta x = 2.5 \times 10^{-4}$ cm, at the infinite far away cold boundary, the temperature of unburned mixture is $T_u =300 K$, and at the left boundary of computing domain, $T = 300.1 K$. The reactions in Table \ref{table: ERH} are all used.

\begin{figure} [ht] 
  \begin{center}
  \centerline{\includegraphics[width=160mm]{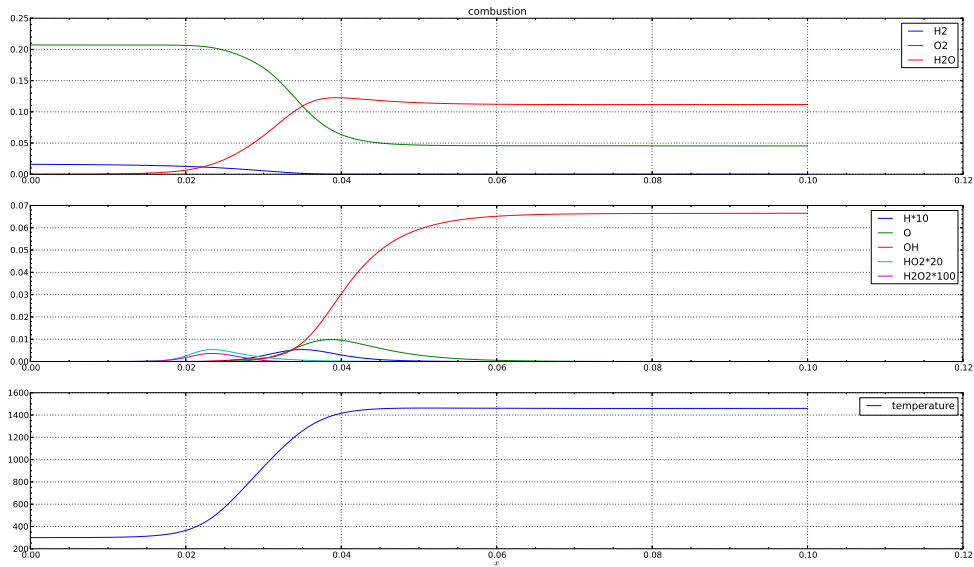}}
  \caption {Flame profile given by the continuation method}
  \small
  The $x$ axes stand for position, in units of center meter (cm). The up two graphs are about mass fraction of all reactive species. In order to make mass fractions of $H$, $HO_2$, and $H_2O_2$ visible in the second graph, their values are magnified by a factor of 10, 20 and 100 respectively. The third is about temperature, with $y$ axis in the unit of Kelvin (K)
  \label{fig: U_new}
  \end{center}
\end{figure}

Applying the continuation method, we get the approximate spatial profile as shown in Fig. \ref{fig: U_new}. The $x$ axes in all three graphs stand for position, in units of center meter (cm). The up two graphs are about mass fraction of all reactive species. In order to make mass fractions of $H$, $HO_2$, and $H_2O_2$ visible in the second graph, their values are magnified by a factor of 10, 20 and 100 respectively. The last one is about temperature, with $y$ axis in the unit of Kelvin (K). We also get the estimated flame speed, $75 cm/s$.

The inflow speed can be improved by a feedback algorithm. Apply the numerical method describe in Section \ref{sec: Numerical Methods} to integrate governing equations (\ref{eqn: conserve rho}) to (\ref{eqn: conserve E}) but turn off all kinds of noise. On the left boundary, the mixture only contains $H_2$, $O_2$ and $N_2$, and $v$ is given; on the right boundary, $p = 1 atm$. Using the profile given by the continuation method as the initial condition, after a short relaxation time, the flame reaches its stable state under deterministic governing equations. The relative flame speed in this reference frame can be decided by tracking the position of a characteristic temperature. Suppose in a period $t_1$, its position moves a distance $L_1$. One can estimate the relative velocity of the flame front with respect to the inflow roughly as $\frac{L_1}{t_1}$. The flame speed by deterministic governing equations is $V_f$ = 75 cm/s - $\frac{L_1}{t_1}$. Reset the inflow speed as
\begin{equation}
v_{new} = v_{old} - a \frac{L_1}{t_1}.
\end{equation}
where $a \in (0, 1]$ is a damping factor. 

\begin{figure} [ht] 
  \begin{center}
  \centerline{\includegraphics[width=160mm]{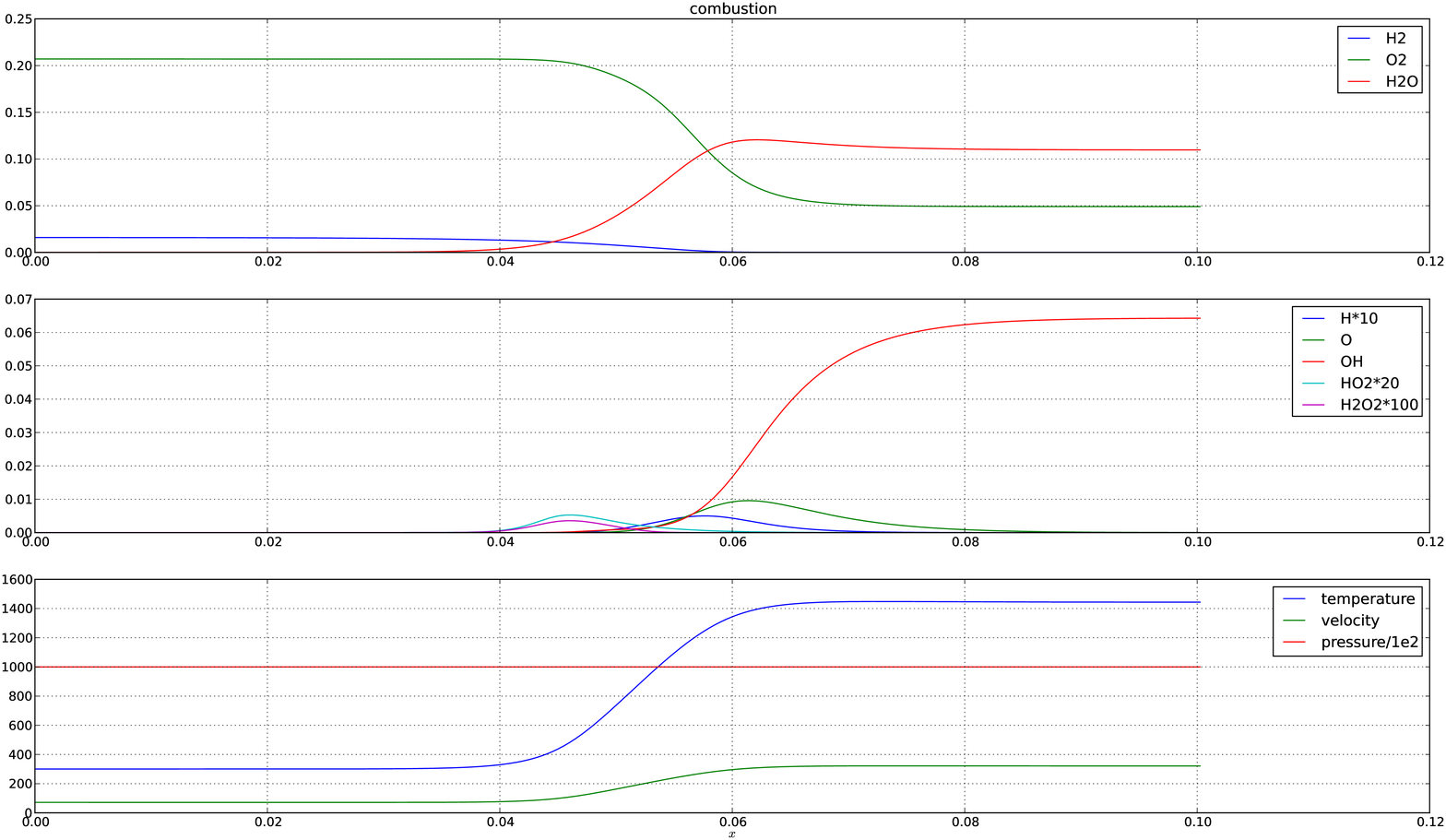}}
  \caption {Flame profile given by time evolving with full deterministic dynamics}
  \small
  The $x$ axes and curves in the up two graphs have the same meaning as in Fig. \ref{fig: U_new}. The last one is about temperature, velocity and pressure with $y$ axis in the units of Kelvin (K), cm/s and Pa respectively. The value of pressure is reduced by a factor of 100.
  \label{fig: U_refine}
  \end{center}
\end{figure}

We choose 800K as our characteristic temperature. By the feedback algorithm, we refine the flame speed as $V_f$ = 71.90 cm/s. The spatial profile is shown in Fig. \ref{fig: U_refine}. The $x$ axes and curves in the up two graphs have the same meaning as in Fig. \ref{fig: U_new}. The last one is about temperature, velocity and pressure with $y$ axis in the units of Kelvin (K), cm/s and Pa respectively. The value of pressure is reduced by a factor of 100. From this profile, we can see that the pressure indeed is almost a constant across the whole domain, which is consistent with the low Mach number approximation in the continuation method.  

\subsection{Influence of noise on hydrogen/air flame speed}
\begin{figure} [ht] 
  \begin{center}
  \centerline{\includegraphics[width=160mm]{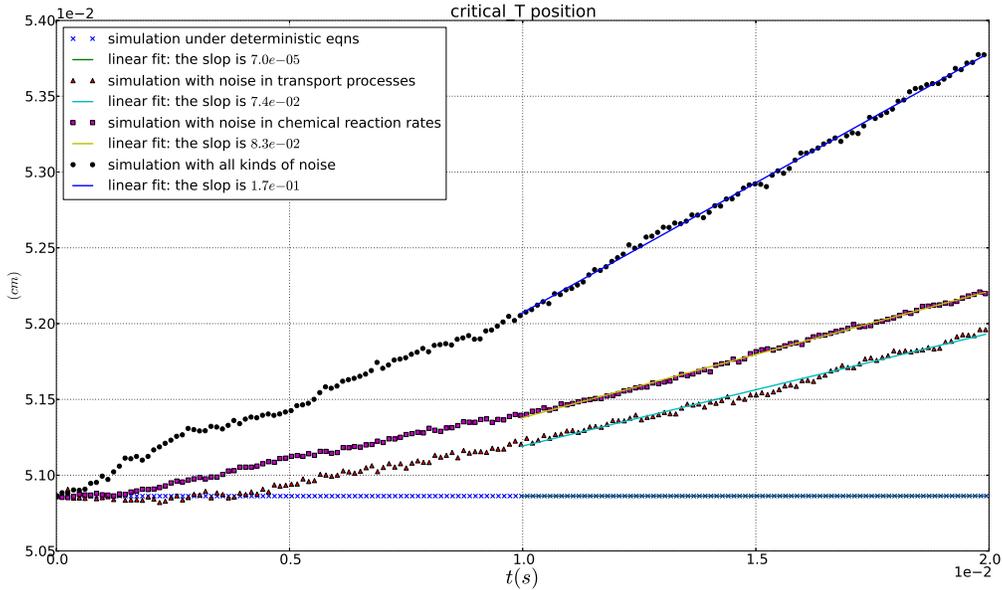}}
  \caption {Position change of Temperature 800 K under different situations}
  \label{fig: T_position}
  \end{center}
\end{figure}

We do numerical experiments under different situations: (1) without any noise; (2) with noise corresponding to transport processes; (3) with noise in chemical reaction rates; (4) with all kinds of noise. We set inflow speed $v$ to be 71.90 cm/s, cross area $10^{-10}$ cm$^2$, and run the simulation for 20 micro seconds. we get Fig. \ref{fig: T_position}. These curves show the position of the chosen temperature, 800K. We do the linear fit to find the mean relative speed when the system roughly reaches its steady state. (1) Without any noise, the relative speed is $7 \times 10^{-5}$ cm/s. (2) When the thermal noise in the transport processes, i.e., the stochastic flux, is turned on, we can find that it is a noisy path. At the beginning, it wanders. It is a period to change from a stable profile under deterministic equations to the one with the stochastic flux. After it reaches its stable state, we estimate the mean relative speed, which is 0.07 cm/s. Compared to the a relative speed $7 \times 10^{-5}$ cm/s under the deterministic dynamics, it is three magnitudes bigger. It is moving to the right side, which suggests that the noise in the transport processes slows down the laminar flame speed. (3) When the statistical noise in reaction rates is turned on, we estimate the mean relative speed as 0.08 cm/s. It also moves to the right side, which suggests that the noise in the reaction rates also decreases the laminar flame speed. (4) When all kinds of noise are turned on, the mean relative speed is 0.17 cm/s, to the right. So overall noise decreases the laminar flame speed 0.24\%.

\begin{figure} [ht] 
	\begin{center}
		\centerline{\includegraphics[width=160mm]{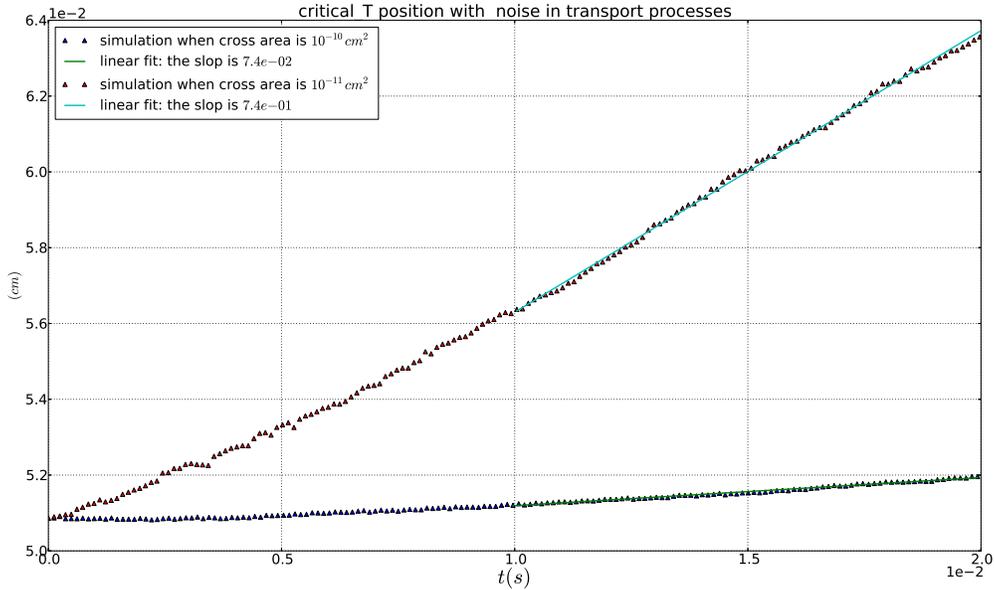}}
		\caption {Influence on propagating front speed with different noise level}
		\label{fig: T_position_2}
	\end{center}
\end{figure}

To understand the relationship between the strength of noise and its effect, we do experiments with different cross areas. With a smaller cross area $10^{-11}$ cm$^2$, the propagating front speed decreases 0.74 $cm/s$ (See Fig. \ref{fig: T_position_2}). This result suggests that noise effect on flame speed is proportional to the variance of noise.

\begin{figure} [ht] 
	\begin{center}
		\centerline{\includegraphics[width=160mm]{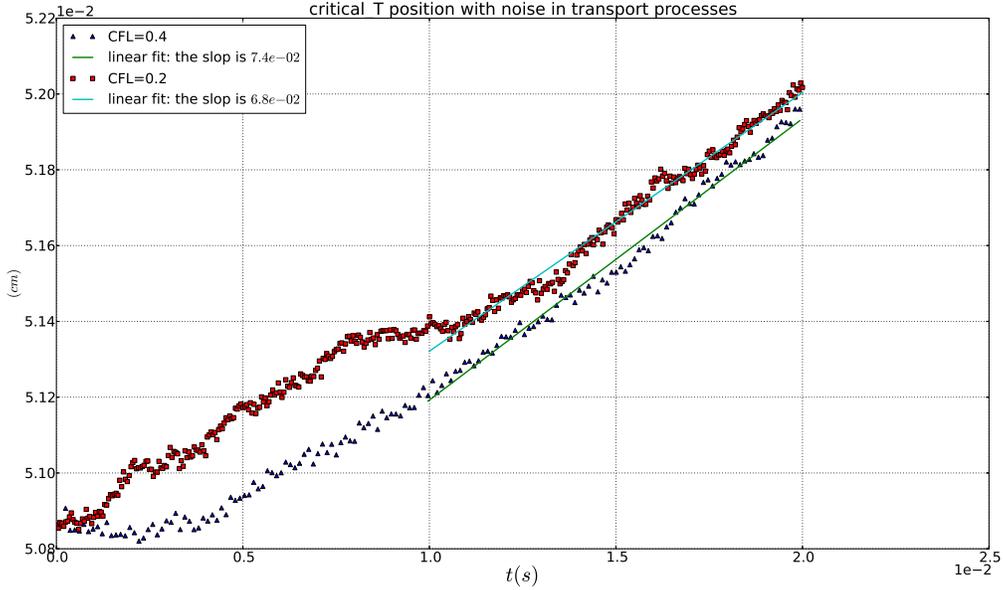}}
		\caption {Influence on propagating front speed with different time steps}
		\label{fig: small_time_step}
	\end{center}
\end{figure}

To study the convergence of this numerical method is too expensive, we only did one more study by a half time step with noise in transport processes. As shown in Fig. \ref{fig: small_time_step}, when $CFL$ number is set to 0.4, under transport noise, the flame speed is decreased by 	0.074 $cm/s$; when $CFL$ number is set to 0.2, under transport noise, the flame speed is decreased by 	0.068 $cm/s$. These results are close. So we believe our result is reliable.

\section{Discussion} \label{sec: discussion}

We developed the code to estimate the noise effect on propagating speeds of premixed laminar flame fronts. We use $H_2$/air as our example. When $\phi = 0.63$, our result shows that noise in transport processes and noise in reaction rates both decrease the flame speed. When all kinds of noise are turned on, with a cross area $10^{-10}$ cm$^2$, the flame speed is decreased by 0.24\%.

When the area of cross section decreases, the phenomena is more significant and our results suggest that noise effect on propagating front speed of a laminar flame is proportional to the variance of noise (i.e. $\epsilon^2$). It is a bit surprising that it is consistent with a simple linear perturbation analysis \cite{thesis} (since the system is so complex).

At last, we point out that noise effect on flames may be more interesting in two and three dimensions.  

\appendix

\section{Deriving the Thermo Fluctuation Terms Corresponding to Transport Processes}\label{appendix 2}
Let us consider a closed system. Suppose the values of a set of variables $\mathbf{x} = \{ x_1, \cdots, x_n \}$ determine its macroscopic state. The equilibrium state is described by 
\begin{equation}
f_{eq}(x_1, \cdots, x_n) = \frac{1}{Z} e^{S(x_1, \cdots, x_n)/k_B} \label{eqn: equilibrium distribution}
\end{equation}
where $Z$ is a normalization factor and $S(x_1, \cdots, x_n)$ is the entropy formally regarded as a function of the exact value of $\mathbf{x}$.
As the system approaches its equilibrium, $x_i$ vary with time. Suppose that the dynamics can be written as,
\begin{equation}
\dot{x}_i = \sum_{j}\gamma_{ij} X_j + y_i \label{eqn: sde}
\end{equation}
where $X_j = \frac{\partial S}{\partial x_j}$ are thermo dynamical forces,  $y_i$ are white noise, and $\{\gamma_{ij}\}$ is the mobility matrix, which is symmetric. In order to make equilibrium distribution satisfy Eq. (\ref{eqn: equilibrium distribution}), the corresponding dynamics should be
\begin{equation}
dx_i = \sum_{j}\gamma_{ij} X_j dt + \sum_{j} \sigma_{ij} dB_j(t),
\end{equation}
where $B_j(t)$ are independent standard Brownian motions and 
\begin{equation}
	\sum_j \sigma_{ij} \sigma_{kj} = 2 k_B \gamma_{ik}.
\end{equation}

According to the laws of thermo dynamics, 
\begin{equation}\label{eqn: thermo law}
	d \epsilon =  T ds + \frac{p}{\rho^2} d\rho + \sum_{i=1}^{N_s}\mu_i dY_i,
\end{equation}
where $s$ is the specific entropy, and $\mu_i$ is the chemical potential of species $i$.
Turn off stochastic terms in Eq. (\ref{eqn: conserve rho}) to Eq. (\ref{eqn: conserve E}), we get the classical deterministic equations of multi-species reactive flows.
With these equations and (\ref{eqn: thermo law}), we get the entropy production rate,
\begin{equation} \label{eqn: s prod rate}
	\sigma = - \frac{\Pi:\nabla\mathbf{v}}{T} - \frac{1}{T^2} \nabla T \cdot \mathbf{Q}' - \sum_{i=1}^{N_s} \mathbf{F}_i \cdot \frac{\nabla_T \mu_i}{T}-\sum_{i=1}^{N_s}\frac{\mu_i \omega_i}{T}.
\end{equation} 
where $\mathbf{Q}' = \mathbf{Q} - \sum\limits_{i=1}^{N_s}h_i \mathbf{F}_i$, and $\nabla_T \mu_i$ is the spatial derivative of $\mu_i$ when holding temperature fixed. 

In the formula Eq. (\ref{eqn: sde}), we take the fluxes $\dot{x}_i$ being $\mathbf{\Pi}$, $\mathbf{F}_i$, and $\mathbf{Q}'$ \cite{Book: LL Fluid}:
\begin{equation}
	\dot{x} \rightarrow \mathbf{\Pi}, \mathbf{F}_i, \mathbf{Q}'.
\end{equation} 
Fluctuations related to $\mathbf{\Pi}, \mathbf{Q}'$ are given in \cite{Book: LL Fluid}. Here we decide the magnitude of $\widetilde{F}_{i,1}$, where $\widetilde{F}_{i,1}$ mean the $\widetilde{\mathbf{F}}_i$ in the $x$ direction. ($\mathbf{F}_i$ in $y$ and $z$ directions are similar and uncorrelated.) Assume we choose $\rho, \mathbf{v}, E$ and $Y_i$, with ${i=1, \cdots, N_s-1}$ as independent variables. By Eq. (\ref{eqn: s prod rate}), the entropy production related to \{$F_{i,1}$\} is
\begin{equation}
  \sigma(\{F_{i,1}\}) = - \frac{1}{T} \sum\limits_{i=1}^{N_s} (\frac{\partial \mu_i}{\partial x})_T F_{i,1} = - \frac{1}{T} \sum\limits_{i=1}^{N_s-1} (\frac{\partial (\mu_i -\mu_{N_s})}{\partial x})_T F_{i,1}.
\end{equation}
Following this, the thermodynamic force in $F_{i,1}$ is $- \frac{1}{T}(\frac{\partial (\mu_i -\mu_{N_s})}{\partial x})_T$. For an ideal gas the chemical potential per mass can be written as,
\begin{equation}
	\mu_i = \frac{R_u T}{W_i}(\mbox{ln} (Y_i \overline{W}/W_i)+ \mbox{ln} p) + g(T), 
\end{equation}
where $g(T)$ is a function only of temperature. 
We have $F_{i,1} = - D_i \frac{\partial Y_i}{\partial x}$, which shows the dynamical equations for $\mathbf{F}_i$ ($i = 1, 2, \cdots, N_s-1$ ) are independent. Recall the approximations we use to get this formula. We have assumed $\nabla p = 0$ and the $N_s^{th}$ species is dominant, which implies $\overline{W} $ can be considered as a constant.  
So we arrive the conclusion, the mobility constant in $F_{i,1}$ is
\begin{equation}
	\frac{- D_i \partial Y_i/ \partial x}{- \frac{1}{T}(\frac{\partial (\mu_i -\mu_{N_s})}{\partial x})_T} = \frac{T D_i}{(\partial \mu_i/\partial Y_i - \partial \mu_{N_s}/\partial Y_i)_T}.
\end{equation}
We have
\begin{equation}
	(\frac{\partial \mu_i}{\partial y_i})_T \approx \frac{R_u T}{W_i} \frac{1}{Y_i},  \qquad
	(\frac{\partial \mu_{N_s}}{\partial y_i})_T \approx - \frac{R_u T}{W_{N_s}} \frac{1}{Y_{N_s}}
\end{equation}
Finally, we get
\begin{equation}
	\widetilde{F}_{i,1} = \sqrt{ \frac{2D_i}{(\frac{N_A}{W_i} \frac{1}{Y_i} + \frac{N_A}{W_{N_s}} \frac{1}{Y_{N_s}})}} W^{(Y_i, 1)},
\end{equation}
where $W^{(Y_i,1)}$ is a white noise random Gaussian vector with uncorrelated components.

Pointed out by \"Ottinger and others \cite{Book: ottinger}, these stochastic terms should be interpreted in the kinetic formula, which can be converted to It$\hat{o}$'s by adding an extra drift term. If this extra drift term is small enough when compared with other drift terms, it is ignored.

\newpage
\section{Elementary Reactions of Hydrogen Flames}
\begin{table}[ht] 
	\caption{Oxidation of $H_2$ \cite{Hydrogen elementary reactions}}
	\centering
	\resizebox{\textwidth}{!}{	\begin{tabular}{c l c r r}
		\hline\hline
		No. & Reaction & B[cm,mol,s] & $\alpha$ & $E_{a}$(kcal/mol) \\
		\hline
		& $H_2 - O_2$ Chain Reactions &&&\\
		(1) & $ H + O_2  \rightleftharpoons O + OH$   & $1.9 \times 10^{14}$ & 0    & 16.44 \\
		(2) & $ O + H_2  \rightleftharpoons H + OH$   & $5.1 \times 10^{04}$ & 2.67 &  6.29 \\
		(3) & $OH + H_2  \rightleftharpoons H + H_2O$ & $2.1 \times 10^{08}$ & 1.51 &  3.43 \\
		(4) & $ O + H_2O \rightleftharpoons OH + OH $ & $3.0 \times 10^{06}$ & 2.02 & 13.40 \\
		& $H_2 - O_2$ Dissociation/Recombination &&&\\
		(5) & $H_2 + M   \rightleftharpoons H + H + M$ & $4.6 \times 10^{19}$ & -1.40 & 104.38 \\ 
		(6) & $O + O + M \rightleftharpoons O_2 + M$   & $6.2 \times 10^{15}$ & -0.50 & 0 \\
		(7) & $O + H + M \rightleftharpoons OH + M $   & $4.7 \times 10^{18}$ & -1.00  & 0 \\
		(8) & $H + OH + M \rightleftharpoons H_2O + M$ & $2.2 \times 10^{22}$ & -2.00  & 0 \\
		& Formation and Consumption of $HO_2$ &&&\\
		(9) &$H + O_2 + M \rightleftharpoons HO_2 + M$ & $6.2 \times 10^{19}$ & -1.42 & 0 \\
		(10) &$HO_2 + H \rightleftharpoons H_2+ O_2 $   & $6.6 \times 10^{13}$ & 0     & 2.13 \\
		(11) &$HO_2 + H \rightleftharpoons OH + OH  $   & $1.7 \times 10^{14}$ & 0     & 0.87 \\
		(12) &$HO_2 + O \rightleftharpoons OH + O_2 $   & $1.7 \times 10^{13}$ & 0     & -0.40 \\ 
		(13) &$HO_2 + OH \rightleftharpoons H_2O + O_2$ & $1.9 \times 10^{16}$ & -1.00 & 0 \\
		& Formation and Consumption of $H_2O_2$ &&&\\
		(14) &$HO_2 + HO_2 \rightleftharpoons H_2O_2 + O_2 $ & $4.2 \times 10^{14}$ & 0     & 11.98 \\
		&                                               & $1.3 \times 10^{11}$ & 0     & -1.629 \\ 
		(15) &$H_2O_2 + M \rightleftharpoons OH + OH + M $   & $1.2 \times 10^{17}$ & 0     & 45.50 \\
		(16) &$H_2O_2 + H \rightleftharpoons H_2O + OH   $   & $1.0 \times 10^{13}$ & 0     & 3.59 \\
		(17) &$H_2O_2 + H \rightleftharpoons H_2 + HO_2  $   & $4.8 \times 10^{13}$ & 0     & 7.95 \\
		(18) &$H_2O_2 + O \rightleftharpoons OH + HO_2   $   & $9.5 \times 10^{06}$ & 2.00  & 3.97  \\
		(19) &$H_2O_2 + OH \rightleftharpoons H_2O + HO_2$   & $1.0 \times 10^{12}$ & 0     & 0 \\
		&                                               & $5.8 \times 10^{14}$ & 0     & 9.56 \\
		
		\hline  
	\end{tabular}}
	
	\label{table: ERH}
\end{table}

\end{document}